\documentclass[12pt]{iopart}

\usepackage{graphicx}
\usepackage{epstopdf}
\usepackage{subfigure}
\usepackage{epsfig}
\usepackage{amssymb,amsfonts}
\usepackage{color}
\begin{document}

\title[Model and parameter dependence of heavy quark energy loss \ldots]{Model and parameter dependence of heavy quark energy loss in a hot and dense medium}

\author{Shanshan Cao $^1$, Guang-You Qin $^{1, 2}$ and Steffen A Bass $^1$}
\address{$^1$ Department of Physics, Duke University, Durham, NC 27708, USA}
\address{$^2$ Department of Physics and Astronomy, Wayne State University, Detroit, MI 48201, USA}
\eads{\mailto{shanshan.cao@duke.edu}, \mailto{qin@phy.duke.edu}, \mailto{bass@phy.duke.edu}}



\begin{abstract}

Within the framework of the Langevin equation, we study the energy loss of heavy quark due to quasi-elastic multiple scatterings in a quark-gluon plasma created by relativistic heavy-ion collisions. We investigate how the initial configuration of the quark-gluon plasma as well as its properties affect the final state spectra and elliptic flow of $D$-meson and non-photonic electron. We find that both the geometric anisotropy of the initial quark-gluon plasma and the flow profiles of the hydrodynamic medium play important roles in the heavy quark energy loss process and the development of elliptic flow. The relative contribution from charm and bottom quarks is found to affect the transverse momentum dependence of the quenching and flow patterns of heavy flavor decay electron; such influence depends on the interaction strength between heavy quark and the medium.

\end{abstract}
\maketitle

\section{Introduction}

It has been well established that a deconfined state of QCD matter, the quark-gluon plasma (QGP), can be created by colliding two heavy nuclei at ultra-relativistic energies, such as those at the Relativistic Heavy-Ion Collider (RHIC) and the Large Hadron Collider (LHC) \cite{Gyulassy:2004zy, Adams:2005dq, Adcox:2004mh, Arsene:2004fa, Back:2004je}. This highly excited form of matter shows properties similar to that of a nearly perfect fluid, such as  strong collective flow \cite{Adams:2003zg, Aamodt:2010pa,ATLAS:2011ah} which has been successfully described by relativistic hydrodynamic calculations \cite{Teaney:2000cw,Huovinen:2001cy,Hirano:2002ds,Gyulassy:2004zy, Nonaka:2006yn,Song:2007fn,Luzum:2008cw,Song:2010mg,Song:2011qa}. 

Another interesting observation is the significant suppression of the production of high transverse momentum hadrons in high energy nucleus-nucleus collisions compared to that in binary-scaled proton-proton collisions at the same energies \cite{Adcox:2001jp, Adler:2002xw, Aamodt:2010jd}. This suppression is commonly understood as the energy loss experienced by high energy partons created in the initial hard collisions as they propagate through the deconfined hot and dense nuclear medium before fragmenting into hadrons \cite{Wang:1991xy, Gyulassy:1993hr, Baier:1996kr, Zakharov:1996fv}. Parton energy loss is thought to originate from a combination of the collisional \cite{Braaten:1991we, Qin:2007rn} and radiative \cite{Wang:1991xy, Gyulassy:1993hr} processes that jets experience when traversing the medium. Sophisticated jet quenching calculations have been performed for single inclusive hadron suppression \cite{Bass:2008rv, Armesto:2009zi, Chen:2010te} as well as di-hadron \cite{Zhang:2007ja, Renk:2008xq, Majumder:2004pt} and photon-hadron correlations \cite{Renk:2006qg, Zhang:2009rn, Qin:2009bk}. Various energy loss schemes have been utilized in those phenomenological applications, and for a detailed comparison between different formalisms, the reader is referred  to Ref. \cite{Armesto:2011ht} and references therein. One of the goals of these calculations is the qualitative extraction of the jet transport parameters in the strongly-interacting QGP medium by comparing the calculations with the measured jet modification data.

The production and the propagation of heavy quarks in relativistic heavy-ion collisions is of similarly great interest as they are expected to improve the constraints on our understanding of jet-medium interaction and provide further insights into the properties of the QGP. There has been observation of significant amount of suppression and elliptic flow for heavy flavor mesons and non-photonic electrons \cite{Adare:2006nq, Abelev:2006db, Adare:2010de,Tlusty:2012ix,Grelli:2012yv,Caffarri:2012wz}. The suppression of heavy flavor decay electrons cannot be explained by the medium-induced radiation alone \cite{Armesto:2005mz, Wicks:2005gt, Qin:2009gw, Younus:2012yi}. Heavy quarks, being massive, suffer medium-induced radiation suppressed in the forward cone compared to that for light partons; this is commonly referred to as the ``dead cone effect" \cite{Dokshitzer:2001zm}. Thus, unlike the light partons whose energy loss is more dominated by medium-induced gluon radiation, the energy loss of heavy quarks is expected to be more dominated by collisional energy loss at low to intermediate transverse momentum \cite{Moore:2004tg, vanHees:2004gq, Mustafa:2004dr, Wicks:2005gt, Gossiaux:2008jv, Qin:2009gw}, though recently it has been stated that the radiation in the backward region may have a larger impact on heavy quark energy loss than expected \cite{Abir:2011jb, Abir:2012pu}.

The large suppression of high momentum heavy quarks and their significant elliptic flow, both deduced from the
measurement of non-photonic electrons, are usually regarded as indication that heavy quarks may thermalize in the QGP \cite{Moore:2004tg, vanHees:2004gq, Adare:2006nq, Abelev:2006db}. However, it was shown in Ref. \cite{Cao:2011et} that the thermalization time of charm quarks might be longer than the lifetime of the QGP phase for reasonable values of the heavy quark diffusion constant. Recent studies seem to indicate that the choice of a variety of medium-related parameters affects heavy quark spectra in relativistic heavy-ion collisions. For instance, Ref. \cite{Gossiaux:2011ea} showed that with the identical transport coefficients, different expansion scenarios can lead to a variation by up to a factor of two for the suppression and elliptic flow of the heavy quark spectra. A similar investigation of the medium flow effect on $D$-meson spectra was performed in Ref. \cite{He:2011qa}. In addition, uncertainties exist in the calculation of initial heavy quark spectra, such as the charm-to-bottom quark ratio \cite{Armesto:2005mz}, which may also affect the non-photonic electron spectra.

In this paper, we investigate how heavy quark spectra in relativistic heavy-ion collisions depend on various ingredients in the phenomenological studies of heavy flavor energy loss, such as the initial production of heavy quarks, the geometry and the flow properties of the hydrodynamic medium, and the coupling strength between heavy quarks and medium. We do not aim for a quantitative description of heavy flavor quenching data due to various uncertainties that we explore here, such as the relative contributions from charm and bottom quarks to the production of non-photonic electrons. The measurements of the suppression and flow for charm and bottom mesons separately will greatly improve this situation and put more constraints on the interpretation of the data. For our purpose, we shall only consider the collisional energy loss experienced by heavy quarks propagating through the QGP medium. Our study should be applicable to heavy quark energy loss in the region from low to intermediate transverse momenta. The contribution from radiative energy loss to heavy quark evolution in medium will be included in an upcoming effort. 

In the limit of multiple quasi-elastic interactions where the energy transfer during each collision is small, the propagation of heavy quarks through a QGP medium can be treated as  Brownian motion. We will follow Ref. \cite{Cao:2011et} and study heavy quark evolution in the medium in the framework of Langevin equation \cite{Svetitsky:1987gq, GolamMustafa:1997id, Moore:2004tg, Akamatsu:2008ge, Gossiaux:2011ea, He:2011qa, Young:2011ug, Alberico:2011zy}. In this framework, one of the crucial parameters is the diffusion coefficient, which has been widely discussed from a microscopic point of view \cite{CasalderreySolana:2006rq, CaronHuot:2007gq, CaronHuot:2008uh, Ding:2011hr}.

The paper is organized as follows. In Sec.\ref{methodology}, we present the methodology utilized for simulating heavy quark evolution in a dynamic medium as produced in relativistic heavy-ion collisions. In Sec.\ref{results}, we present the numerical results of the nuclear modification factor and elliptic flow for heavy quarks, heavy flavor mesons, and non-photonic electrons, where their dependence on various inputs are discussed in details. A summary and outlook is presented in Sec.\ref{summary}.

\section{Methodology}
\label{methodology}

In this work, we only consider the collisional energy loss experienced by heavy quarks propagating in the QGP medium. The multiple quasi-elastic scatterings of heavy quarks inside a thermalized medium can be treated as brownian motion and be described by the Langevin equation:
\begin{equation}
\frac{d\vec{p}}{dt}=-\eta_D(p)\vec{p}+\vec{\xi}.
\end{equation}
In principle, the noise term $\vec{\xi}$ may depend on the momentum of heavy quark. In this work we do not consider such dependence and the noise term is assumed to satisfy the following correlation relation:
\begin{equation}
\langle\xi^i(t)\xi^j(t')\rangle=\kappa\delta^{ij}\delta(t-t').
\end{equation}
Such treatment is sufficient in the non-relativistic limit. In the relativistic case, the detailed momentum dependence relies on the internal structure and the properties of the medium. The investigation of such dependence and its effect on heavy quark energy loss and flow will be left for a future effort. 

To implement a numerical simulation of the momentum evolution of heavy quarks with the Langevin equation, the Ito discretization is adopted,
\begin{eqnarray}
\label{noise}
&&\vec{p}(t+\Delta t)=\vec{p}(t)-\vec{d}_{Ito}(\vec{p}(t))\Delta t+\vec{\xi}\Delta t,  \nonumber\\
&&\langle\xi^i(t)\xi^j(t-n\Delta t)\rangle=\frac{\kappa}{\Delta t}\delta^{ij}\delta^{0n},
\end{eqnarray}
where the force due to the drag is given by
\begin{equation}
\vec{d}_{Ito}(\vec{p})=\eta_D(p)\vec{p}.
\end{equation}
Assuming the energy transfer is small, it can be shown that the fluctuation-dissipation relation applies, which indicates:
\begin{equation}
\eta_D(p)=\frac{\kappa}{2TE}.
\end{equation}
Furthermore, according to Eq.(\ref{noise}), Gaussian noise with width
$\Gamma=\sqrt{\left. \kappa \right / \Delta t}$
will be used to generate the random momentum kicks in our calculation. The diffusion coefficient is related to the drag term via:
\begin{equation}
D=\frac{T}{M\eta_D(0)}=\frac{2T^2}{\kappa}.
\end{equation}

The detailed information about the produced highly excited medium in relativistic heavy-ion collisions is essential to heavy quark energy loss and the development of elliptic flow. Throughout our calculation, the QGP medium is described by a fully (3+1)-D relativistic ideal hydrodynamic model developed in Ref. \cite{Nonaka:2006yn}. In our work, two different initial condition models, a Glauber \cite{Glauber:1970jm, Miller:2007ri} as well as a KLN-CGC \cite{Kharzeev:2004if} model,  are utilized to describe the initial energy distribution of the medium before the hydrodynamic evolution. In the Glauber model, the collision between two nuclei is viewed in term of the individual interactions between the constituent nucleons; while in the KLN-CGC model, the unintegrated gluon distributions inside the two colliding nuclei are used to determine the production and distribution of the initial  gluons. These two initial state models provide the energy/entropy density profiles with different spatial anisotropies in the transverse plane, a larger eccentricity for the KLN-CGC than the Glauber. The comparison between the two will allow for a study of the sensitivity of heavy-quark observables to the initial spatial make-up of the system. We will focus on mid-central Au-Au collisions at RHIC with a center-of-mass energy $\sqrt{s_\mathrm{NN}}$=200~GeV per nucleon pair and take the impact parameter $b=6.5$~fm throughout the calculation.

In the hydrodynamic simulation of the QCD medium, the initial time $\tau_0$ is chosen to be 0.6~fm/c. Up to now, little knowledge has been attained for the pre-equilibrium evolution and thermalization of the system. Therefore, for heavy quark motion prior to the QGP formation, we treat it as free-streaming. Such treatment should be good approximation as the time of the pre-equilibrium stage is short compared to the total life time of the QGP (about 10fm/c). Any heavy quark leaving the QGP will stream freely as well.

The hydrodynamic simulation provides us with the time-evolution and the spatial distribution of temperature and flow velocity for the thermalized QGP medium. In such a dynamic medium, heavy quark evolution is treated as follows: for every Langevin time step we boost the heavy quark to the local rest frame of the fluid cell through which it propagates. The Langevin approach is then applied to obtain the momentum evolution of heavy quark. After that we boost it back to the global computation frame.

Since the production of heavy quarks is dominated by the processes with large transverse momentum transfer, perturbative QCD is applied to calculate the initial momentum distribution of heavy quarks prior to their propagation through the QGP medium. We fit the leading-order perturbative QCD calculation with a power-law distribution \cite{Moore:2004tg}, and sample the initial transverse momentum of heavy quarks according to the following parametrization:
\begin{equation}
\label{distribution}
\frac{dN}{d^2 p_\mathrm{T}} \propto \frac{1}{(p^2_\mathrm{T}+\Lambda^2)^\alpha},
\end{equation}
where $\alpha=3.9$ and $\Lambda=2.1$ for charm quarks, and $\alpha=4.9$ and $\Lambda=7.5$ for bottom quarks. In this work, we focus on the energy loss of heavy quarks at mid-rapidity and therefore assign no initial longitudinal momentum to heavy quarks. We have checked that the introduction of initial longitudinal momenta that are uniform around the mid-rapidity region ($-1<\eta<1$) does not affect our final transverse momentum spectra and does not affect the systematics we are about to explore. The relative normalization (ratio) of charm and bottom quarks is not fixed, but rather serves as a free parameter in our simulation. Later we will investigate the effect of this normalization on the quenching and the elliptic flow of heavy flavor decay electrons.

The initial spatial distribution of heavy quarks in the transverse plane is sampled according to the distribution of binary collisions as calculated from a Monte-Carlo Glauber model. With the spatial and momentum initialization of heavy quarks, we are able to simulate their time evolution inside the QGP medium in the framework of Langevin equation as described above. After passing through the medium, their fragmentation into heavy flavor mesons and the subsequent decay into electrons are simulated via Pythia 6.4 \cite{Sjostrand:2006za}. By default, the fragmentation process is calculated with the Lund symmetric fragmentation function that is modified by the Bowler spaceÐtime picture of string evolution \cite{Bowler:1981sb} for heavy quark. And the hadronic and the subsequent semi-leptonic processes are combined for the decay of charm/bottom hadrons in which all possible channels are taken into account. Details about the implementation are discussed in the manual above.

In the end, the final state particles in the mid-rapidity region ($-1<\eta<1$) are selected and their momentum distribution are utilized to calculate the elliptic flow coefficient $v_2$ and the nuclear modification factor $R_\mathrm{AA}$ as follows:
\begin{eqnarray}
&& v_2(p_\mathrm{T})=\langle \cos(2\phi)\rangle=\left\langle\frac{p_x^2-p_y^2}{p_x^2+p_y^2}\right\rangle, \nonumber\\
&& R_\mathrm{AA}=\frac{\left({dN}/{dp_\mathrm{T}}\right)_{\textnormal{fin}}}{\left({dN}/{dp_\mathrm{T}}\right)_{\textnormal{init}}}.
\end{eqnarray}
Note that when heavy quarks are directly analyzed, the denominator and the numerator of $R_\mathrm{AA}$ are the initial heavy quark distribution and the distribution of those surviving from the energy loss and passing through the medium. When analyzing heavy flavor mesons or electrons, the denominator represents the spectra of the corresponding particles fragmented/decayed directly from the initial heavy quarks, while the numerator represents those produced from the heavy quarks after transporting through the QGP medium.

\section{Numerical Results}
\label{results}

In this section, we present the numerical results of the simulation of heavy quark evolution in the QGP medium produced by Au+Au collisions at the RHIC energy. We investigate the impact of various parameter choices of the calculation on the final spectra and elliptic flow of heavy quarks, heavy mesons and their decay electrons.

\subsection{Charm Quark Energy Loss and Flow}

The energy loss of heavy quarks and the development of elliptic flow crucially depend on the geometrical 
shape and dynamical evolution of the thermalized QGP medium that heavy quarks traverse. The total energy loss of heavy quarks is mostly controlled by the overall magnitude of the energy density of the medium, while the elliptic flow is more sensitive to the geometry of the medium as it characterizes the anisotropy of the final transverse momentum spectra. In typical non-central nucleus-nucleus collisions, the overlap region of the two nuclei is anisotropic in the transverse plane, thus resulting in the anisotropy of the produced hot and dense medium. Due to the different pressure gradients in different directions, different radial flows are built up during the hydrodynamic evolution of the thermalized QGP.

In such an anisotropic dynamical medium, there exist two factors affecting the anisotropy of heavy quark energy loss: the different path lengths and the different flow profiles experienced by the heavy quarks traveling in different directions. Longer paths will be traveled through by heavy quarks moving in the out-of-plane ($y$) direction than in the in-plane ($x$) direction, where the reaction plane is defined to be spanned by the impact parameter and the beam axis directions. Thus in absence of collective flow for the medium, heavy quarks, after passing through such anisotropic medium, would have larger momentum in the  $x$ direction than in the $y$ direction, $\langle p_x^2 \rangle > \langle p_y^2 \rangle$, resulting in a positive elliptic flow. In addition, the collective flow of the medium also contributes positively to heavy quark elliptic flow since the push of the radial flow is more prominent in the $x$ direction. Therefore, the total elliptic flow developed during the propagation of heavy quarks in such an anisotropic hydrodynamic medium is due to a combination of these two factors.

\begin{figure}[tb]
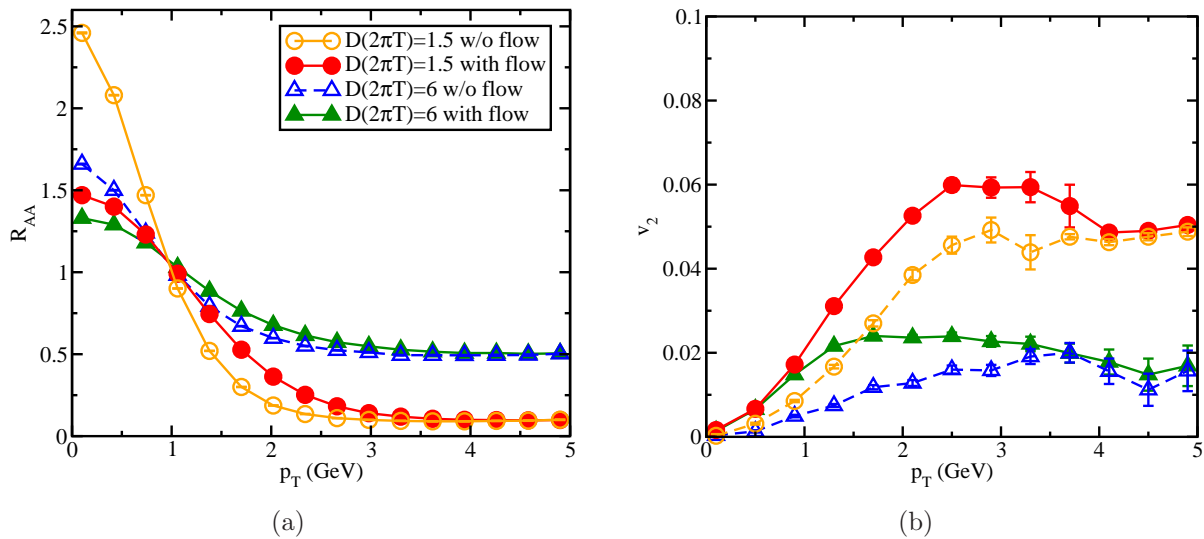

\subfigure[]{\label{floweffectRAA}\epsfig{file=RAA_flowEffect_c_G.eps, width=0.48\textwidth, clip=}}
\hspace{0.04\textwidth}
\subfigure[]{\label{floweffectv2}\epsfig{file=v2_flowEffect_c_G.eps, width=0.48\textwidth, clip=}}
\vspace{-1pc}
 \caption{(Color online) A comparison between the influence of QGP media with and without collective flow on (a) $R_\mathrm{AA}$ and (b) $v_2$ of charm quarks. Both media are generated with the Glauber initial condition.}
 \label{floweffect}
\end{figure}

We can separate these two effects in the simulation by turning on or off the coupling of the collective flow of the thermalized medium to the evolving  heavy quarks. The decoupling from the collective flow can be accomplished by not boosting the heavy quarks into the respective rest frame of the fluid cell for the Langevin evolution. The comparison between the heavy quark evolution with and without coupling to the collective flow  is shown in Fig.\ref{floweffect}, where the left plot shows the nuclear modification factor $R_\mathrm{AA}$ and the right plot shows the elliptic flow $v_2$ of the charm quarks as a function of the transverse momentum. We show results for two different values of diffusion coefficient $D = 1.5/(2\pi T)$ and $D = 6/(2\pi T)$.

The effect of the collective flow of the medium on the heavy quark energy loss can be clearly seen from the plot of the nuclear modification factor $R_\mathrm{AA}$ (Fig.\ref{floweffect}). It is negligible at high transverse momenta, and becomes observable at intermediate transverse momentum regime. Due to the push by the radial flow, heavy quarks are less suppressed (i.e. have a larger $R_\mathrm{AA}$) at larger transverse momenta, since the radial flow effectively transports low momentum heavy quarks to larger transverse momenta. Similar effects stemming from the elliptic flow of the medium are observed for the heavy quark elliptic flow coefficient $v_2$. At low transverse momenta, the collective flow of the medium presents a significant influence on the charm quark $v_2$. At high transverse momenta, the collective flow effect is small, thus the development of charm $v_2$ is dominated by the geometric anisotropy of the medium. The dominance of the medium collective flow at low transverse momentum for $v_2$ might indicate that low transverse momentum charm quarks are more likely to lose a significant amount  of their momenta and therefore thermalize in the medium, and thus flow more like the thermalized medium.

A closer observation suggests that with a decrease of the diffusion coefficient, i.e., an increase of the coupling strength, the influence of the geometric asymmetry becomes more dominant. For instance, Fig.\ref{floweffect} reveals that for $D=6/(2\pi T)$, the geometric asymmetry of the medium contributes to only approximately half of the charm quark $v_2$ at the peak value (around $p_\mathrm{T}=1.5$~GeV). However,  for $D=1.5/(2\pi T)$, such contribution increases to more than 80\% at the corresponding peak value (around $p_\mathrm{T}=3$~GeV). Note that such increase of the geometric contribution is not unlimited. With further reduction of the diffusion coefficient ($D < 1.5/(2\pi T)$), i.e., a larger coupling between heavy quarks and the medium, the energy loss of charm quarks will be so intense that all of them will be captured by the medium. In that limit, charm quarks thermalize with the medium during the QGP lifetime \cite{Cao:2011et}, and therefore, their $v_2$ will entirely follow the collective flow of the medium. In our simulation, the choice of $D \sim 1.5/(2\pi T)$ provides the largest elliptic flow for the final heavy quarks.

\begin{figure}[tb]
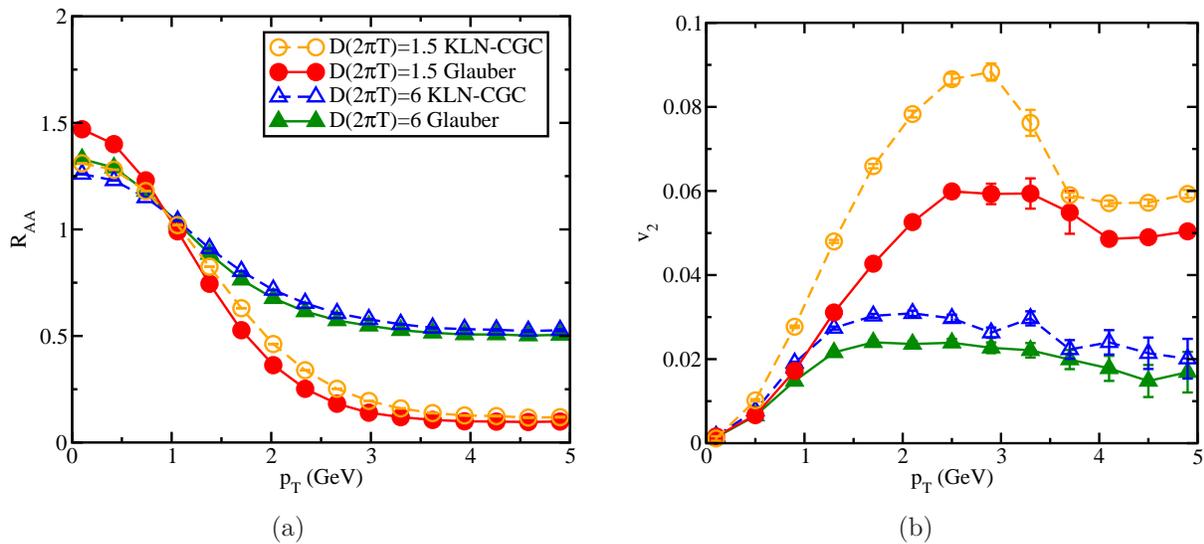

\subfigure[]{\label{hydroeffectRAA}\epsfig{file=RAA_hydroEffect_c.eps, width=0.48\textwidth, clip=}}
\hspace{0.04\textwidth}
\subfigure[]{\label{hydroeffectv2}\epsfig{file=v2_hydroEffect_c.eps, width=0.48\textwidth, clip=}}
\vspace{-1pc}
 \caption{(Color online) A comparison between the influence of QGP media with the Glauber and the KLN-CGC initial conditions on (a) $R_\mathrm{AA}$ and (b) $v_2$ of charm quarks.}
 \label{hydroeffect}
\end{figure}

We may further investigate the  effect of the spatial medium distribution on the heavy quark energy loss  and the development of heavy quark elliptic flow by utilizing different initial conditions for the hydrodynamic simulation of the QGP. Two different initial condition models are widely used for the initialization of the energy density distribution prior to the hydrodynamic evolution: the Glauber model \cite{Glauber:1970jm, Miller:2007ri} 
and KLN parametrization of the Color Glass Condensate (CGC) model  \cite{Kharzeev:2004if}. These two models provide initial energy density profiles with different anisotropies in the transverse plane. In particular, the KLN-CGC model exhibits a larger eccentricity $\epsilon_2 = \langle y^2-x^2 \rangle / \langle y^2 + x^2 \rangle$ than the Glauber model, which will manifest itself in larger elliptic flow coefficients for the heavy quarks.

The comparison between these two initial condition models is shown in Fig.\ref{hydroeffect}, where the left frame of the figure shows the nuclear modification factor $R_\mathrm{AA}$ and right shows the elliptic flow $v_2$. As expected, a significantly larger elliptic flow is observed for the charm quarks traveling through the hydrodynamic medium with the KLN-CGC initial condition than those with the Glauber initial condition. As indicated by Fig.\ref{hydroeffect}, the difference can be as large as 20\% for $D = 6/(2\pi T)$ and 40\% for $D =1.5/(2\pi T)$. We also observe that while $v_2$ is sensitive to the choice of the initial condition, the nuclear modification factor $R_\mathrm{AA}$ is not significantly affected by the choice of these two hydro initial conditions. This is due to $R_\mathrm{AA}$ being controlled by the overall normalization of the density profile in the hydrodynamic evolution which has been tuned to describe the properties of bulk matter, such as the $\pi$ and $K$ spectra.

\subsection{$D$ Mesons and Heavy Decay Electrons}

In the above discussion, we have focused on the effects of initial conditions and medium parameters  on 
heavy quark energy loss and the development of heavy quark  elliptic flow. Now we investigate the corresponding sensitivities of heavy flavor mesons and their decay electrons. Since the KLN-CGC initial condition provides a larger eccentricity for the initial energy density profile and thus produces a larger elliptic flow of heavy quarks during their medium evolution, we use it for the remainder of our analysis. This is merely to obtain the largest possible values of the final elliptic flow, since most of the previous calculations seem to under-predict the elliptic flow data of non-photonic electrons once the model parameters have been tuned to describe the measured nuclear modification factor.

\begin{figure}[tb]
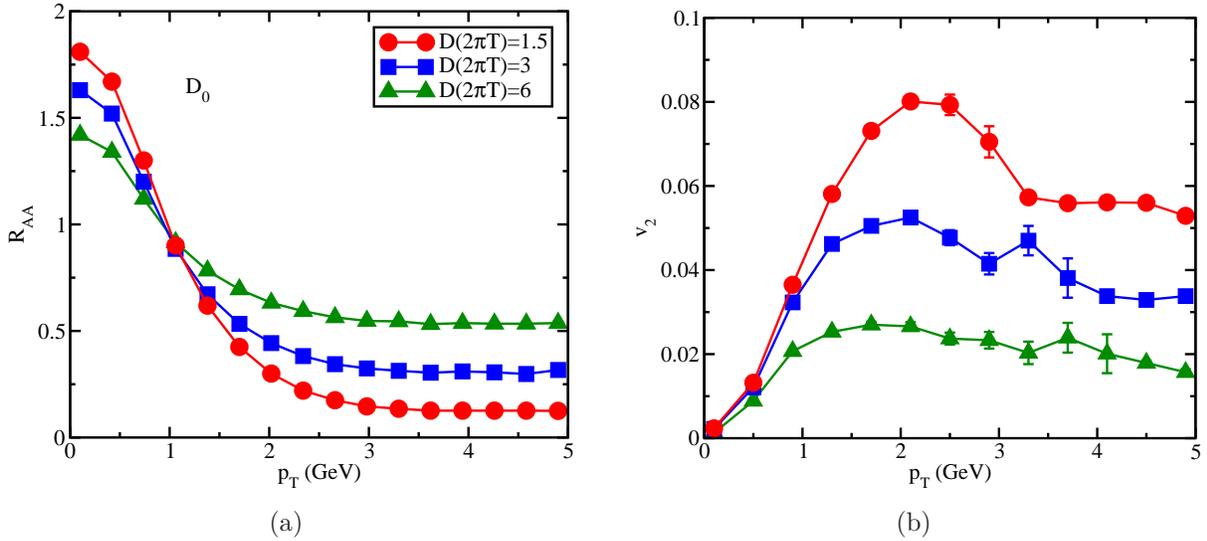

\subfigure[]{\epsfig{file=RAA_D_C.eps, width=0.48\textwidth, clip=}}
\hspace{0.04\textwidth}
\subfigure[]{\epsfig{file=v2_D_C.eps, width=0.48\textwidth, clip=}}
\vspace{-1pc}
 \caption{(Color online) (a) $R_\mathrm{AA}$ and (b) $v_2$ of $D_0$ mesons. The QGP medium is generated with the KLN-CGC initial condition.}
 \label{Dmeson}
\end{figure}

\begin{figure}[tb]
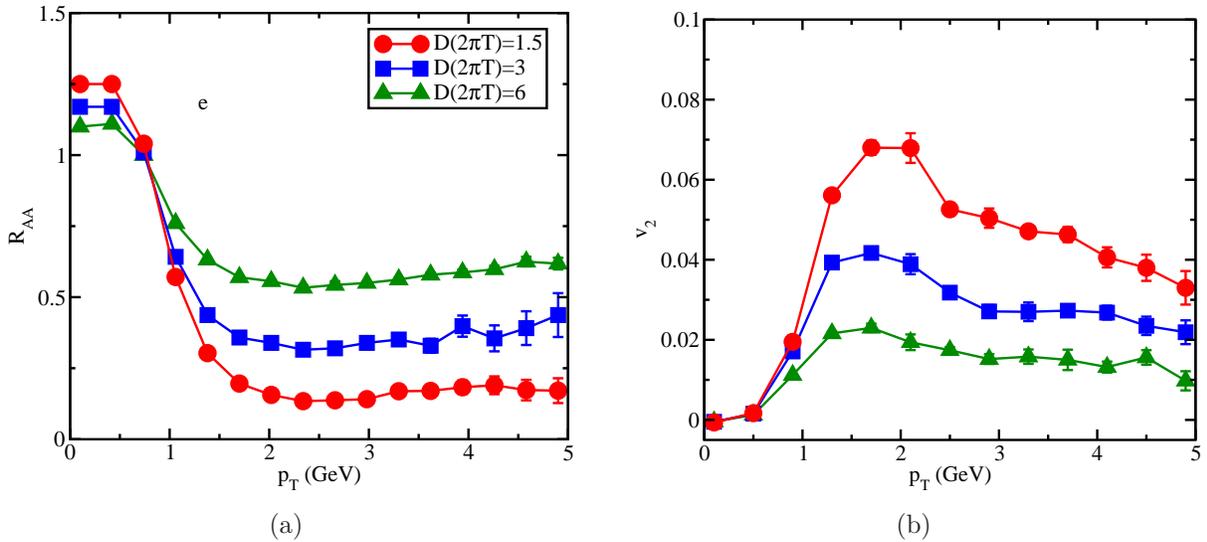

 \subfigure[]{\epsfig{file=RAA_e_C.eps, width=0.48\textwidth, clip=}}
 \hspace{0.04\textwidth}
 \subfigure[]{\epsfig{file=v2_e_C.eps, width=0.48\textwidth, clip=}}
 \vspace{-1pc}
 \caption{(Color online) (a) $R_\mathrm{AA}$ and (b) $v_2$ of electrons decayed from charm quarks. The QGP medium is generated with the KLN-CGC initial condition.}
 \label{efromc}
\end{figure}

Figures \ref{Dmeson} and \ref{efromc} display the numerical results of the nuclear modification factor $R_\mathrm{AA}$ and elliptic flow $v_2$ for $D$ mesons and $D$-decay electrons. Three different values of diffusion coefficients are used for comparison $D = 1.5/(2\pi T)$, $D = 3/(2\pi T)$, and $D = 6/(2\pi T)$. We observe that the transverse momentum dependence of $R_\mathrm{AA}$ and $v_2$ are similar to that for charm quarks as shown in the previous figures.

\begin{figure}[tb]
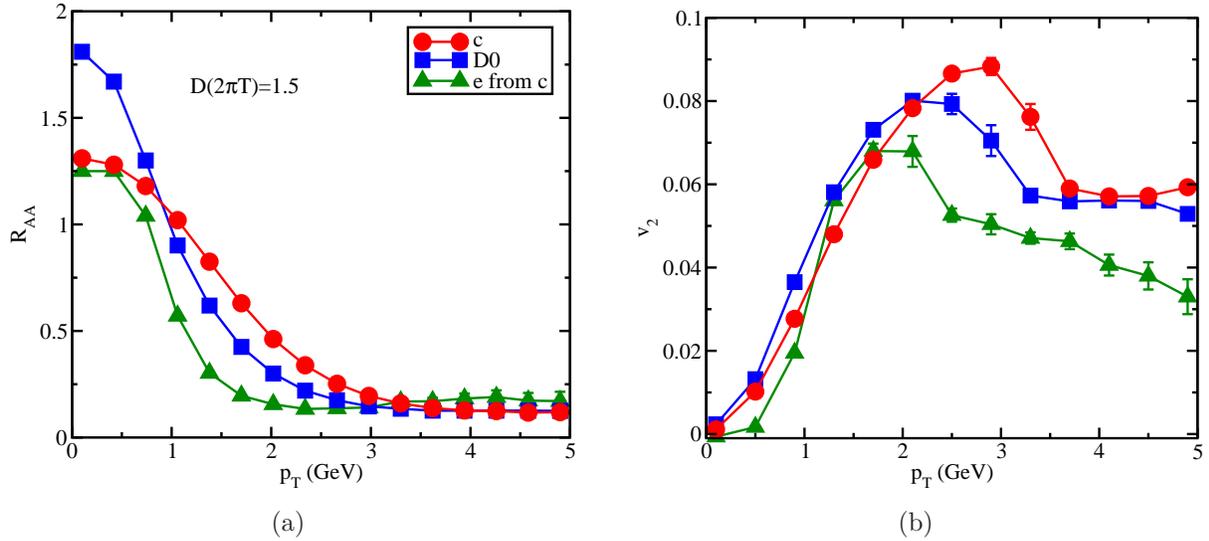

 \subfigure[]{\epsfig{file=RAA_D1-5.eps, width=0.48\textwidth, clip=}}
 \hspace{0.04\textwidth}
 \subfigure[]{\epsfig{file=v2_D1-5.eps, width=0.48\textwidth, clip=}}
 \vspace{-1pc}
 \caption{(Color online) A comparison of (a) $R_\mathrm{AA}$ and (b) $v_2$ between charm quarks, $D_0$ mesons and electrons. The QGP medium is generated with the KLN-CGC initial condition.}
 \label{compare_cDe}
\end{figure}

Figure \ref{compare_cDe} provides a more direct comparison of the nuclear modification of $R_\mathrm{AA}$ and elliptic flow $v_2$ between charm quarks, $D$ meson and $D$-decay electrons. Here the result is shown for a diffusion coefficient of $D=1.5/(2\pi T)$; other values of the diffusion coefficient display similar systematics. 
One observes that all the nuclear modification factors $R_\mathrm{AA}$ curves decrease with increasing transverse momentum, and saturate above $p_\mathrm{T} \sim 3-4$~GeV, while the elliptic flow curves first increase, reach a peak and then decrease (and saturate). The peak values of $v_2$ is the largest for charm quarks and the smallest for electrons. This  decrease of the momentum anisotropy can be understood as a result of additional randomization of the momentum space during the hadronization and decay processes. \footnote{This order is obtained with the fragmentation mechanism only for heavy flavor hadronization and may change with the introduction of the coalescence mechanism. We shall discuss this effect in detail in a coming work.} Moreover, the values of the transverse momentum at which $v_2$ reaches the peak value shift to lower regimes from charm quarks to $D$ mesons and to $D$-decay electrons. This pattern is caused by the decrease of the momentum from the parent particles to the daughter particles during heavy quark fragmentation and heavy flavor meson decay processes.

\begin{figure}[tb]
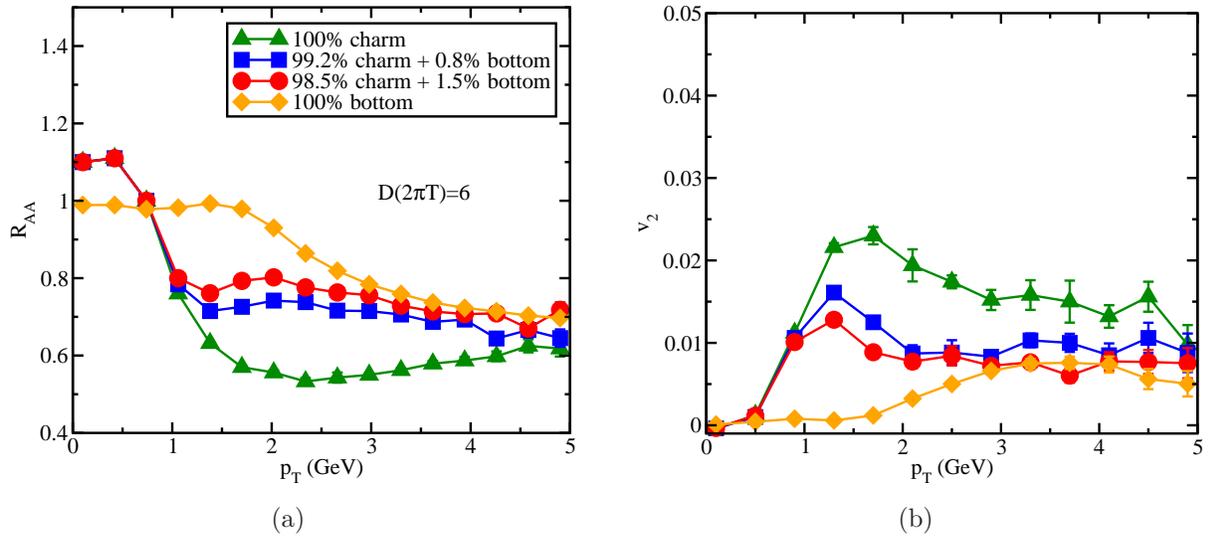

 \subfigure[]{\label{cbD6RAA}\epsfig{file=RAA_cbD6_e_C.eps, width=0.48\textwidth, clip=}}
 \hspace{0.04\textwidth}
 \subfigure[]{\label{cbD6v2}\epsfig{file=v2_cbD6_e_C.eps, width=0.48\textwidth, clip=}}
 \vspace{-1pc}
 \caption{(Color online) A comparison of (a) $R_\mathrm{AA}$ and (b) $v_2$ of non-photonic electrons between different initial charm/bottom ratios. Set $D=6/(2\pi T)$ and the QGP medium is generated with the KLN-CGC initial condition.}
 \label{cbD6}
\end{figure}

\begin{figure}[tb]
 \subfigure[]{\label{cbD1-5RAA}\epsfig{file=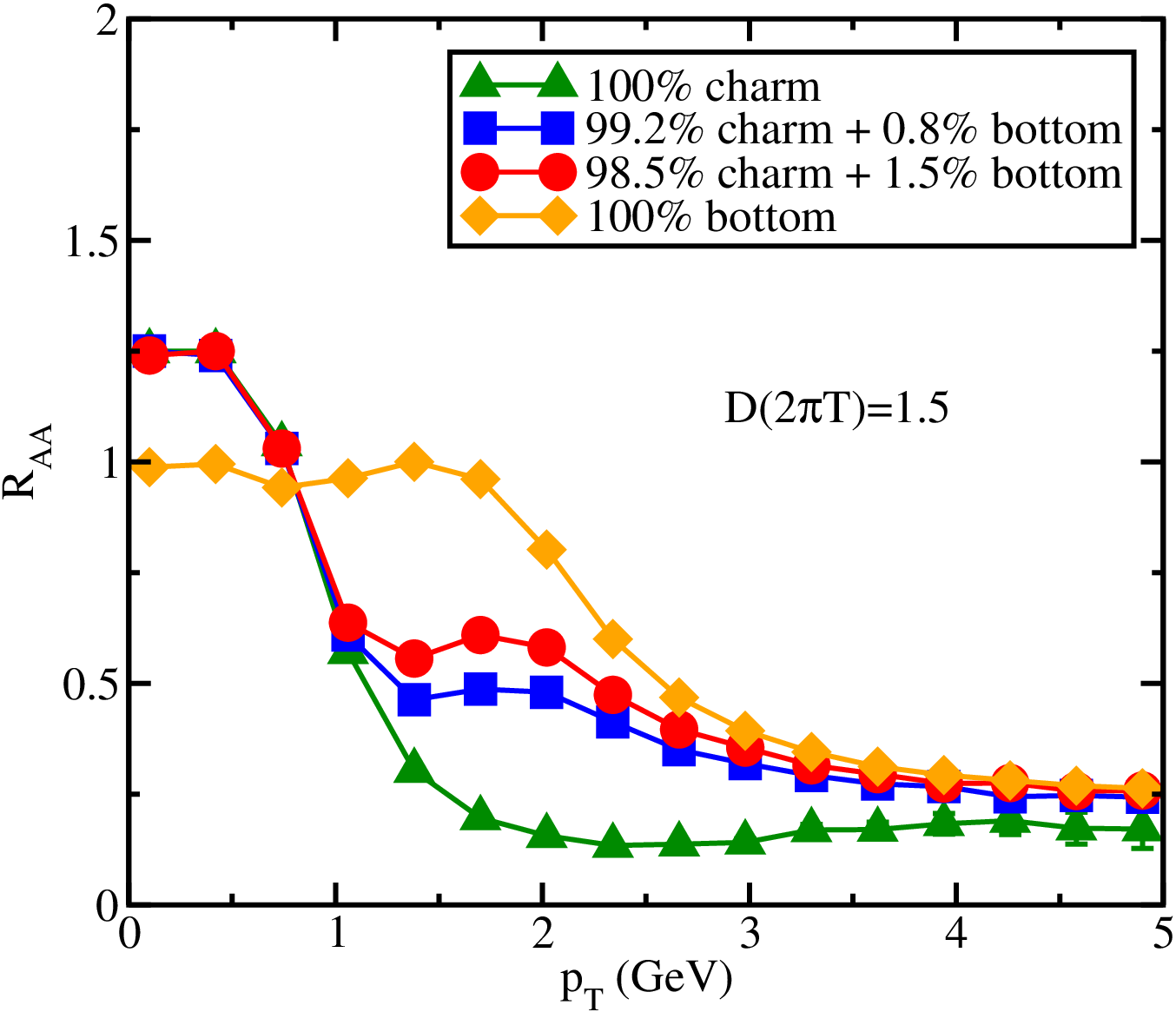, width=0.48\textwidth, clip=}}
 \hspace{0.04\textwidth}
 \subfigure[]{\label{cbD1-5v2}\epsfig{file=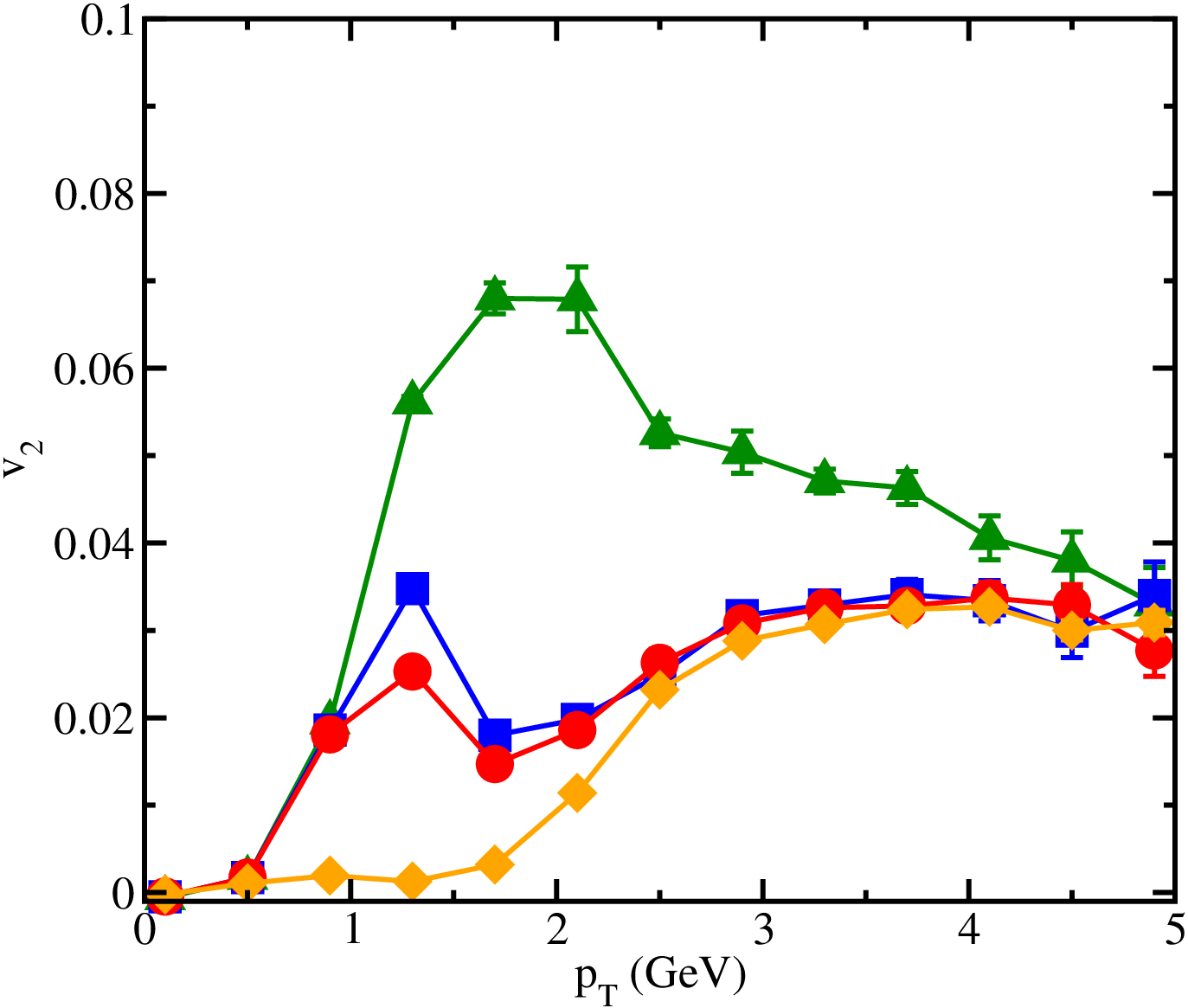, width=0.48\textwidth, clip=}}
 \vspace{-1pc}
 \caption{(Color online) A comparison of (a) $R_\mathrm{AA}$ and (b) $v_2$ of non-photonic electrons between different initial charm/bottom ratios. Set $D=1.5/(2\pi T)$ and the QGP medium is generated with the KLN-CGC initial condition.}
 \label{cbD1-5}
\end{figure}

For the heavy flavor decay electron spectra, another important factor is the relative contributions from charm vs. bottom quarks. Since charm and bottom quarks have different masses, they are produced with different initial transverse momentum distributions, and experience different energy loss and coupling to the collective flow in medium. This manifests itself in different $R_\mathrm{AA}$ and $v_2$ systematics for $D$ and $B$ mesons respectively and subsequently translates into different behavior for their respective decay electrons. The electrons at lower transverse momentum are dominated by charm quark decay, while in the high transverse momentum regime bottom quark dominates as the source of these electrons. Since there are multiple uncertainties affecting the relative normalization of charm and bottom quark production, for example the scale dependence in the perturbative QCD calculation of initial heavy quark production \cite{Armesto:2005mz}, we treat the ratio of charm and bottom quarks as a free parameter for our calculation, and investigate how 
the variation of this ratio affects the final non-photonic electron distributions.

The results are shown in Fig.\ref{cbD6} and  Fig.\ref{cbD1-5} for two different values of diffusion coefficients, $D=1.5/(2\pi T)$ and $D=6/(2\pi T)$, respectively. We compare four different initializations here -- pure charm, pure bottom, and two mixtures of charm and bottom quarks: 99.2\% charm quarks with 0.8\% bottom quarks, and 98.5\% charm quarks with 1.5\% bottom quarks. As shown in \cite{Armesto:2005mz}, the bottom quark contribution to the electron spectra may start dominating over charm quark contribution at transverse momentum as low as 3~GeV or as high as 9~GeV. Our two hybrid mixtures of charm and bottom quarks have about a factor of 2 difference in their ratio, representing an estimate of the uncertainties due to our limited control of the proton-proton baseline.

One observes from these two figures that the nuclear modification factor $R_\mathrm{AA}$ and the elliptic flow $v_2$ of heavy flavor decay electrons are very different for the pure charm vs. pure bottom scenario. Bottom quarks are less suppressed than charm quarks at high transverse momenta, thus less enhancement is observed at low transverse momenta in the $R_\mathrm{AA}$. The magnitude of the elliptic flow coefficient $v_2$ is much smaller for electrons from bottom decay than from charm decay, again due to the reduced energy loss experienced by the bottom quarks. In addition, we observe a difference in the transverse momentum dependence:  while the elliptic flow coefficient $v_2$ of electrons from charm decays has a peak value at intermediate transverse momentum, that for bottom decays increases monotonically with increasing transverse momentum (and then saturates). This result from the fact that charm quarks contribute mostly to the production of low transverse momentum electrons while bottom quarks contribute mostly to high transverse momentum electrons.

Due to the different behavior of charm vs. bottom decay electrons, the electrons from a mixture of charm and bottom decays exhibit a very rich structure. Both $R_\mathrm{AA}$ and $v_2$ trend similar to the pure charm initialization at  low transverse momentum and converge to the values of the pure bottom quark scenario at high transverse momenta. In the intermediate transverse momentum region where the transition from charm dominance to bottom dominance in the origin of the decay electrons takes place, a non-monotonic transverse momentum dependence of $R_\mathrm{AA}$ and $v_2$ is observed: a dip-peak structure for $R_\mathrm{AA}$ and a peak-dip structure for $v_2$. Such a non-monotonic behavior is more prominent for the smaller value of the diffusion coefficient $D=1.5/(2\pi T)$ (Fig.\ref{cbD1-5}), since a smaller value of the diffusion coefficient increases the interaction with the medium and thus the energy loss of charm quarks and their elliptic flow, while such an enhancement is far less for bottom quarks due to their larger mass. Current experimental results seem not able to determine whether such a peak-dip structure is present or not in the non-photonic electron  elliptic flow $v_2$ due to large experimental error bars. Further improvement of the measurement of the  detailed transverse momentum dependence of non-photonic electrons would be helpful for the determination of the diffusion coefficient and therefore the coupling strength between heavy quarks and the QGP medium.

Another important effect seen in Fig.\ref{cbD6} and Fig.\ref{cbD1-5} is the significant sensitivity of heavy flavor decay electron $v_2$ to the initial charm-to-bottom quark ratio. For instance, a  $0.7\%$ difference in
the mixing ratio between charm and bottom quarks in our simulation leads to a variation of approximately 25\% in $v_2$ for a diffusion coefficient of $D=6/(2\pi T)$ and over 30\% for $D=1.5/(2\pi T)$. As has been discussed earlier, significant uncertainties regarding the initial heavy quark spectra are still present in our current phenomenological calculations, and thus provide a sizable uncertainty for the prediction of the quenching and 
elliptic flow of non-photonic electrons.

\section{Summary and outlook}
\label{summary}

In this work, we have studied the energy loss of heavy quarks in a hot and dense medium produced in relativistic heavy-ion collisions. The Langevin approach is utilized to simulate the heavy quark evolution in the medium due to quasi-elastic multiple scatterings. Numerical results are presented for both the nuclear modification factor and the elliptic flow of heavy quarks, heavy flavor mesons and their corresponding  non-photonic decay electrons. We have investigated in details how the final nuclear modification factor and elliptic flow are affected by various components of the model, such as the geometry and the collective flow of the hydrodynamic medium, the initial production ratio of charm to bottom quarks and the coupling strength between the heavy quarks and the medium.

We have focused on two particular properties of the medium that affect the heavy quark energy loss -- its geometric anisotropy and its collective flow. It is found that the geometric anisotropy dominates the final heavy quark distributions in the high transverse momentum region, while the collective flow of the medium dominates the low momentum region. The impact of the QGP geometry on the heavy quark energy loss has been explored by comparing the Glauber and the KLN-CGC initialization of the hydrodynamic medium. We found that while a similar nuclear modification factor $R_\mathrm{AA}$ is observed for both initial condition models, a significantly higher heavy quark elliptic flow $v_2$ is found  for the KLN-CGC model. We have further investigated the sensitivity of the spectra and elliptic flow of non-photonic electrons to the relative contributions from charm and bottom quarks. It is found that a less than $1\%$ difference in the initial charm-to-bottom ratio can lead to more than $30\%$ variation of the non-photonic electron spectra. Therefore, narrowing down these uncertainties is essential for a better understanding of the interaction dynamics between heavy quarks and the QGP medium.

Though we have not preformed a detailed quantitative comparison to the experimental measurements, most of our results can be tested with future measurements. For instance, the spectra of the heavy flavor decay electrons show a dip-peak structure of $R_\mathrm{AA}$ and a peak-dip structure of $v_2$ (see Fig.\ref{cbD1-5}) for large values of the quark-medium coupling; such structures would disappear if the coupling becomes weaker (Fig.\ref{cbD6}). Due to the large error bars of the data in the high transverse momentum region, current experimental observations can not yet distinguish between these two scenarios. Future high precision measurements that can be compared to this prediction will help determine the coupling strength between heavy quarks and the QGP medium.

While our study constitutes an important step towards the quantitative understanding of the interaction dynamics of heavy quarks in a hot and dense medium, it can be further improved in several directions, which we leave for future work. Here, we have utilized the Langevin approach for the simulation of heavy quark evolution as affected by multiple quasi-elastic scatterings in the QGP medium, in which many details regarding the microscopic structures of the interaction between heavy quarks and the medium are encoded in the diffusion coefficient. However, the heavy quark scattering cross section in the non-perturbative regime remains largely unknown, resulting in different predictions of this transport coefficient -- it varies from less than $D=2/(2\pi T)$ based on a lattice QCD calculation \cite{Ding:2011hr} to over $D=7/(2\pi T)$ according to the $T$-matrix method \cite{He:2011qa}, and therefore leading to large uncertainties in the prediction of the final state spectra. Another interesting effect on the heavy quark evolution in medium is the radiative energy loss induced by multiple scatterings. This effect has been discussed in other frameworks like the Boltzmann equation \cite{Gossiaux:2010yx}, but is still absent in the current Langevin algorithm and may be important, especially in the high transverse momentum regime. In addition, the coalescence mechanism for heavy flavor hadronization may have non-negligible impact at low energies and therefore affect the final state observables.  In a follow-up work, we shall address some of these questions, in particular how to incorporate gluon radiation for heavy quark energy loss  and the coalescence mechanism for its hadronization into our current Langevin framework. This will set the basis  for a direct comparison to experimental data from both RHIC and LHC.

\section*{Acknowledgments}

We are grateful to the Open Science Grid for providing computing resources that were used in this work. We acknowledge many useful discussions with Berndt M\"uller.  This work was supported by U.S. department of Energy grant DE-FG02-05ER41367 and the NSF under grant No. PHY-1207918.

\section*{References}
\bibliography{SCrefs}

\end{document}